\documentclass{article}
\usepackage{spconf,amsmath,graphicx}
\usepackage{diagbox}
\usepackage{multirow}
\usepackage{color}
\usepackage{graphicx}
\usepackage{cases}
\usepackage{subfig}
\usepackage{booktabs}
\usepackage{boldline}
\usepackage{amsfonts}
\usepackage{cite}
\usepackage{balance}
\usepackage{soul}
\usepackage{threeparttable,tablefootnote}
\usepackage[english]{babel}
\usepackage[colorlinks,linkcolor=red,anchorcolor=blue,citecolor=green,CJKbookmarks=True]{hyperref}       
\usepackage{url}            

\title{Speaker and Direction Inferred Dual-channel Speech Separation}
\name{Chenxing Li$^{1,2}$, Jiaming Xu$^{1,2\dagger}$\thanks{$\dagger$ Corresponding author}, Nima Mesgarani$^{3}$, Bo Xu$^{1,2\dagger}$}
\address{
$^1$Institute of Automation, Chinese Academy of Sciences, Beijing, China  \\
$^2$University of Chinese Academy of Sciences, Beijing, China \\
$^3$Columbia University, New York, NY, USA
}

\begin{document}
\ninept
\maketitle
\begin{abstract}
Most speech separation methods, trying to separate all channel sources simultaneously, are still far from having enough generalization capabilities for real scenarios where the number of input sounds is usually uncertain and even dynamic. In this work, we employ ideas from auditory attention with two ears and propose a speaker and direction inferred speech separation network (dubbed SDNet) to solve the cocktail party problem. Specifically, our SDNet first parses out the respective perceptual representations with their speaker and direction characteristics from the mixture of the scene in a sequential manner. Then, the perceptual representations are utilized to attend to each corresponding speech. Our model generates more precise perceptual representations with the help of spatial features and successfully deals with the problem of the unknown number of sources and the selection of outputs. The experiments on standard fully-overlapped speech separation benchmarks, WSJ0-2mix, WSJ0-3mix, and WSJ0-2\&3mix, show the effectiveness, and our method achieves SDR improvements of 25.31 dB, 17.26 dB, and 21.56 dB under anechoic settings. Our codes will be released at~\url{https://github.com/aispeech-lab/SDNet}.


\end{abstract}
\section{Introduction}
\label{sec:intro}

In many environments, the auditory scene is composed of several concurrent speech streams with their spectral features overlapping both in space and time. Human auditory system exhibits a remarkable ability to parse these complex scenes. However, background noise, overlapping speech, and reverberation damage the quality and degrade the performance of speech recognition.

Recently, some researchers attempt to alleviate the problem and pay extensive attention to neural network-based speech separation. In the single-channel-based separation task, many methods have achieved state-of-the-art (SOTA) performance, such as frequency domain-based DPCL \cite{hershey2016deep}, DANet \cite{chen2017deep}, PIT \cite{kolbaek2017multitalker}, Chimera++ \cite{wang2018alternative}, CBLDNN-GAT \cite{li2018cbldnn}, SPNet \cite{wang2019deep}, Deep CASA \cite{liu2019divide} and time domain-based TasNet \cite{luo2019conv}, FurcaPa \cite{shi2019deep}, DPRNN\cite{luo2020dual}. These methods design the model structure from different perspectives and follow different training strategies, where the factors affecting performances are investigated in depth. However, these methods meet several challenges: an unknown number of sources in the mixture, permutation problem, and selection from multiple outputs.

In order to deal with the situation that the number of sources in mixed speech is unknown, paper \cite{higuchi2017deep} incorporates DPCL into the masking-based beamforming and performs separation. OR-PIT \cite{takahashi2019recursive} separates only one speaker from a mixture at a time, and the residual signal is sent to the separation model for the recursion to separate the next speaker. An iteration termination criterion is proposed to identify the number of speakers accurately. A speaker-inferred model \cite{shi2019ones} uses the Seq2Seq-based method \cite{bahdanau2015neural,yang2018sgm} to infer speakers. Speaker information is also appended to the output.
Auxiliary autoencoding PIT \cite{luo2020separating} is proposed to further improve the performance across various numbers of speakers.

Speaker-aware-based networks \cite{wang2018deep,delcroix2018single, wang2019voicefilter,xu2020spex,ge2020spex+} try to deal with the problem of permutation and selection from outputs.
These methods are interested in recovering a single target speaker while reducing noise and the effect of interfering speakers. The reference speech from the target speaker should be given in advance.

In addition to the single-channel-based methods, multi-channel-based methods can extract additional direction features to further improve the performance, and some methods are proposed to solve the problems of permutation and output selection. Similar to the speaker-aware-based networks, Li et al. \cite{li2019direction} use fixed beamformers to transfer the multi-channel mixture into single-channel signals. An attention network is designed to identify the direction of the target speaker and combine the beamformed signals. SpeakerBeam \cite{delcroix2018single} is then applied to separating the enhanced signal. The direction-aware-based method \cite{gu2020temporal} focuses on the target source in a specific direction by using a time domain-based network.

\begin{figure*}[t]
  \centering
  \includegraphics[width=0.91\linewidth]{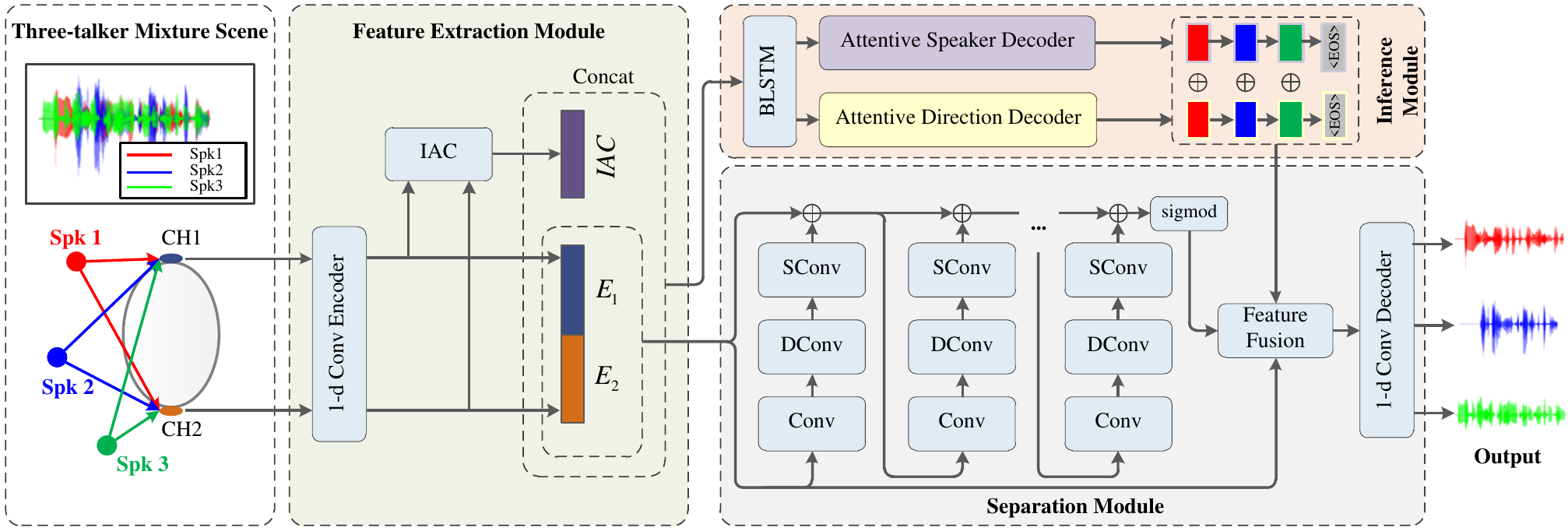}
  \caption{The model architecture of SDNet. In the separation module, SConv and DConv represent the depth-wise separable convolution \cite{chollet2017xception}.}
  \label{fig:model}
\end{figure*}

PIT-based methods \cite{kolbaek2017multitalker,li2018cbldnn,wang2019deep,luo2019conv,shi2019deep,luo2020dual} need prior knowledge about the number of speakers and meet the permutation problem. These methods have some shortcomings in real environments. In the existing methods account for identifying the number of outputs, \cite{takahashi2019recursive} requires iterative operations, which increases system complexity. \cite{higuchi2017deep,takahashi2019recursive} still can not solve the problem of the selection of outputs. Speaker-aware-based methods need to know the target speaker in advance. The speech of other speakers cannot be separated. Besides, in single-channel-based methods, speakers with similar pitch are difficult to be separated. By extending it to multi-channel-based methods, an additional direction feature can be acquired by the network.

In real environments, a source signal has a unique speaker and direction information. We propose a speaker and direction-inferred dual-channel speech separation network (SDNet), which can infer speaker and direction information first and use them as cues to separate speech. Our contributions are listed as follows: (1) We expand single-channel to multi-channel time domain-based separation based on \cite{shi2019ones}. Spectral and spatial features are fully utilized. (2) Instead of manually extracting channel differences in \cite{wang2018integrating, gu2019end}, the channel differences are extracted by the network and can be optimized end-to-end. (3) This network can simultaneously infer speaker and direction information, and the information is fused as a source mask for separation. By dynamically estimating the number of source masks, the network can cope with the problem of the unknown number of outputs. (4) After separation, speaker and direction information are appended to the separated speech. This information can be used in subsequent tasks. The network can deal with the problem of the selection of outputs.

Scale-invariant signal-to-noise ratio (SISNR) \cite{luo2019conv} and signal-to-distortion ratio (SDR) \cite{vincent2006performance} improvements are used to evaluate the performance. Experimental results show that SDNet can effectively perform separation both on the anechoic and reverberant settings.

\section{System Overview}

The illustration of our model is shown in Fig.~\ref{fig:model}. Our network is composed of three components: (1) The feature extraction module processes features from each channel and extracts the channel differences; (2) The inference module parses out the speakers and directions from the mixture and generates source masks; (3) The separation network processes features and integrates the source masks to generate the separated outputs.

\subsection{Feature extraction module}

\subsubsection{convolutional encoder}

The encoder transforms the mixture waveforms into an intermediate feature space. In detail, the input segment is transformed into the representation by using a one-dimensional (1D) convolutional layer.

\begin{equation}
  E_i = \mathit{Conv1d(CH_i)},~~i=1,2,
\label{encoder1}
\end{equation}
where $E$ indicates the encoder, and $\mathit{CH_i}$ represents the waveform of $i$-th channel.

\subsubsection{inter-channel attention correlation}

Compared with the single-channel-based models, dual-channel-based models can use both spatial and spectral information. This is conducive to the improvement of performance. The time difference between the channels can be obtained by end-to-end training or manually-designed features \cite{wang2018integrating, gu2019end}. We directly calculate the correlation among the channels and integrate it into the network as an additional feature. In detail, similar to the self-attention \cite{vaswani2017attention}, we calculate the attention correlation between channels. By setting channel 1 as the reference, the inter-channel attention correlation (IAC) is as follows:

\begin{equation}
\mathit{IAC}=\mathit{\textup{softmax}(E_1E_{2}^{T})},
  \label{IAC}
\end{equation}
where $\mathit{E_1}$ and $\mathit{E_2}$ represent the output of encoder 1 and encoder 2, respectively. Channel differences may contribute to the inference module. The feature extraction module has two different outputs, and the outputs are:
\begin{equation}
\begin{aligned}
\mathit{F}=\mathit{[E_1, E_2]},~~~~\mathit{F_o}=\mathit{[IAC, E_1, E_2]},
\label{FEout}
\end{aligned}
\end{equation}
where $F_o$ is sent to the inference module, and $F$ is fed into the separation module.

\subsection{Inference module}

In the inference module, the Seq2Seq-based mechanism \cite{bahdanau2015neural,yang2018sgm} is applied to inferring the speakers and directions in a sequence manner. First, the features are mapped into high-level vectors by using stacked bi-directional long short term memory (BLSTM) layers. The specific equation is:

\begin{equation}
  \mathit{h} =\mathit{BLSTM(F_o)},
  \label{Pencoder}
\end{equation}
where $h$ is the hidden state, and $F_o$ represents the input feature of the inference module.

We use two independent decoding networks to infer the speakers and the directions, respectively. Considering not all speech features make contributions to infer the speakers and directions equally at each step, the attention mechanisms are utilized to produce context vectors by focusing on different portions of the sequence and aggregating the hidden representations. Two attentive decoding networks have similar procedures. Here we formulate the speaker-inferred decoder as follows:

\begin{equation}
\begin{aligned}
\mathit{\alpha_{ti}}&=\mathit{\textup{softmax}(tanh(W_1s_{t-1}+U_1h_i))}, 
\label{Pattention}
\end{aligned}
\end{equation}
\begin{equation}
\begin{aligned}
\mathit{c_t}&=\sum^T_{i=1}\mathit{\alpha_{ti}h_i},
\end{aligned}
\end{equation}
where $W_1$, $U_1$ are weights, and $s_{t-1}$ is the hidden state of the decoder at time-step $t-1$. $c_t$ is the context vector at time-step $t$.

For the decoding networks, a global embedding strategy is introduced to alleviate the problem of exposure bias~\cite{yang2018sgm}, and the embedding feature at $t$ is calculated as follows:
\begin{equation}
\begin{aligned}
e^{a}_{t}&=\sum^N_{j=1}{y^{j}_{t-1}e_j},
\label{Pembedding}
\end{aligned}
\end{equation}
\begin{equation}
\begin{aligned}
\mathit{g}&=\mathit{\textup{sigmoid}(W_2e_t+U_2e^{a}_{t})}, 
\end{aligned}
\end{equation}
\begin{equation}
\begin{aligned}
\mathit{e_{s_t}}&=\mathit{g\odot e_{t} + (1-g)\odot e^a_{t}},
\end{aligned}
\end{equation}
where $N$ is the number of speakers. $y^{j}_{t-1}$ is the $j$-th element of $y_{t-1}$ and $e_j$ is the embedding vector of the $j$-th speaker. $e^{a}_{t}$ denotes the weighted average embedding at time $t$. $W_2$ and $U_2$ are weight matrices. $e_{s_t}$ represents the speaker embedding at time $t$. $\odot$ denotes element-wise multiplication. The hidden state $s_t$ of the decoder at time-step $t$ is computed as follows:
\begin{equation}
\mathit{s_t}=\mathit{\textup{LSTM}(s_{t-1}, [e_{s_t};c_{t}])}.
\label{PLSTM}
\end{equation}
The final output is calculated as:
\begin{equation}
\begin{aligned}
\mathit{y_t}&=\mathit{\textup{softmax}(W_3f(W_4s_t + W_5c_t))},
\label{Py}
\end{aligned}
\end{equation}
where $W_3$, $W_4$, and $W_5$ are weights. $y_t$ represents the inferred probability distribution of the inferred speaker at time-step $t$. In each time step, rather than selecting the final output $y_t$, the speaker embedding, $e_{s_t}$, is selected as the speaker mask. When the inferred $y_t$ corresponds to an $\mathit{<}$EOS${>}$ (End-of-Sequence), the decoding process is stopped.

The inference process of direction mask, $e_{d_t}$, is the same as the equations above. The source mask can be obtained as follows:
\begin{equation}
\mathit{sm_t}=\mathit{e_{s_t}+e_{d_t}},
\label{Pmask}
\end{equation}
where $sm_t$ means the $t$-th source mask inferred by the inference module. In the inference module, these two attentive decoders run simultaneously. If one decoder infers an $<$EOS$>$, the two decoders are stopped. In the test, the beam search algorithm \cite{wiseman2016sequence} is applied to finding the top-ranked inference.

\subsection{Separation module}
Temporal convolutional networks (TCN) \cite{bai2018empirical} effectively memorize long-term dependencies. Dilation rate \cite{bai2018empirical} is used to continuously expand the receptive field. The separation module is the same as the separation module in TasNet \cite{luo2019conv}. In detail, the streamline of the separation module consists of four convolutional blocks. In each block, for expanding receptive fields, dilated convolutional operations are repeat $R$ times with 1,2,4,..., and $2^{R-1}$ dilation rates. A sigmoid activation then scales the output.

To generate separated outputs, the decoding process is the inverse process of the encoding layer. It decodes the feature representation to speech samples. Specifically, we use 1D transposed convolution to implement the decoding process:

\begin{equation}
\begin{aligned}
\mathit{Z_i} &= \mathit{F\odot TCN_o\odot sm_i,~~i=1,...,n,} \\
\mathit{D(Z)_i} &= \mathit{TransposedConv(Z_i),~~i=1,...,n},
\label{Separation}
\end{aligned}
\end{equation}
where $TCN_o$ denotes the output of TCN layers. $Z_i$ represents the high-level feature representatives of $i$-th inferred source. $n$ is the number of source masks infered in this mixture. $D(\cdot)_i$ represents the $i$-th separated output.

\subsection{Loss function}

End-to-end training is performed, and three kinds of loss are adopted: raw-waveform-based SiSNR separation loss, cross-entropy-based speaker-inferred loss, and cross-entropy-based direction-inferred loss. The detailed loss function is formulated as:

\begin{equation}
\mathit{\mathcal{L}=-\mathcal{L}_{\mathit{SiSNR-SS}}+\lambda\times(\mathcal{L}_{\mathit{CE-Spk}}+\mathcal{L}_{\mathit{CE-Dir}}}),
\end{equation}
where $\lambda$ is a hyper-parameter. For the inference module, speaker indexes act as the speaker labels, which are 101 in this experiment. 37 directions are chosen as the direction labels, which are distributed from 0 degrees to 180 degrees with a 5-degree interval. The labels of direction are generated during the data simulation. Meanwhile, $<$BOS$>$ (Begin-of-Sequence) and $<$EOS$>$ are also added to the speaker and direction label sets. For each sample, $<$BOS$>$ is placed at the top of the labels, and $<$EOS$>$ is placed at the end. $<$BOS$>$ means that the network starts to infer. $<$EOS$>$ is used for the network to determine the end of decoding.

\begin{table}[!t]
        \caption{\label{tab:results1} {\it The effect of SNet-time in single-channel anechoic datasets and comparison of different methods on SDR improvement (dB).}}
        \centerline{
          \begin{tabular}{c c c c}
            \toprule[1pt]
            {System}  &{WSJ0-2mix} &{WSJ0-3mix} &{WSJ0-2\&3mix}\\
            \hline
            \hline
            SNet-time                &$12.35$    &$9.87$ & $10.81$ \\
            \hline
              SNet \cite{shi2019ones}  &$7.52$   &$5.14$  & $7.05$ \\
              DPCL++  \cite{isik2016single}  &$10.3$   &$7.1$  & $8.8$ \\
              uPIT-BLSTM \cite{kolbaek2017multitalker}  &$10.0$   &$7.7$  & $8.9$ \\
              TasNet  \cite{luo2019conv}                  &$15.0$    &$12.8$   & $-$ \\
              OR-PIT  \cite{takahashi2019recursive}       &$15.0$   &$12.9$   & $-$\\
             \bottomrule[1pt]
          \end{tabular}
        }
\end{table}

\begin{table*}[t]
        \caption{\label{tab:results2} {\it The effect of different configurations on dual-channel datasets and comparisons on SISNR and SDR improvement (dB).}}
        \centerline{
          \begin{tabular}{c c c c c c c c c c}
            \toprule[1pt]
            \multirow{2}*{System}  &\multirow{2}*{Domain}  &\multirow{2}*{Data Type} &\multicolumn{2}{c}{WSJ0-2mix} & \multicolumn{2}{c}{WSJ0-3mix} & \multicolumn{2}{c}{WSJ0-2\&3mix}\\
            \cline{4-9}
            & & &{SISNRi} & {SDRi}  &{SISNRi}  &{SDRi} & {SISNRi}  &{SDRi}\\
            \hline
            \hline
            SNet-2ch           & Freq.  &Anechoic   &$14.62$        &$14.25$        &$10.31$         &$10.03$         & $11.12$ & $11.02$ \\
            SNet-time-2ch      & Time   &Anechoic   &$20.88$        &$20.61$        &$14.32$         &$14.11$         & $17.43$ & $17.31$ \\
            SNet-time-2ch+IAC  & Time   &Anechoic   &$21.13$        &$20.89$        &$15.41$        &$15.02$          & $18.65$ & $18.11$ \\
            SDNet              & Time   &Anechoic   &$25.71$ &$25.31$ &$17.46$  &$17.26$  &$21.92$ &$21.56$ \\
            \hline
            TasNet-2ch                              &Time  &Anechoic    &$25.21$  &$25.08$     &$17.31$     &$17.06$     & $-$ & $-$ \\
            \bottomrule[0.8pt]
            SNet-2ch         & Freq.  &Reverberant    &$7.32$    &$7.28$       &$5.53$  &$5.15$    & $6.53$ & $6.33$ \\
            SNet-time-2ch      & Time  &Reverberant   &$8.43$    &$8.35$       &$6.62$  &$6.41$    & $7.32$ & $7.30$\\
            SNet-time-2ch+IAC  & Time  &Reverberant   &$8.76$    &$8.59$       &$6.93$  &$6.86$    & $7.88$ & $7.64$\\
            SDNet          & Time  &Reverberant       &$10.57$     &$10.64$       &$8.49$    &$8.55$      &$9.91$ &$9.08$ \\
            \hline
            TasNet-2ch          &Time  &Reverberant  &$10.78$  &$10.83$     &$9.08$     &$9.32$     & $-$ & $-$ \\
            \bottomrule[1pt]
          \end{tabular}
        }
\end{table*}

\section{Experiments}

\subsection{Experimental setup}

The proposed methods are evaluated on 8k Hz single and dual-channel WSJ0-2mix, WSJ0-3mix, and WSJ0-2\&3mix datasets \cite{hershey2016deep}. For both single-channel and stereo datasets, WSJ0-2mix and WSJ0-3mix contain 30 hours of training data, 10 hours of development data, and 5 hours of test data. The mixing signal-to-noise ratio, pairs, dataset partition are exactly coincident with paper \cite{hershey2016deep}. WSJ0-2\&3mix is the union of WSJ0-2mix and WSJ0-3mix. Anechoic and reverberant stereo datasets are generated by convolving the clean speech with the room impulse responses \cite{allen1979image,lehmann2008prediction}. For reverberant datasets, the reverberation time is uniformly sampled from 40 ms to 200 ms. We place 2 microphones at the center of the room. The distance between microphones is 10 cm. Sound sources are randomly placed in the room. The training set and the test set contain 101 and 18 speakers, respectively. The speakers in the test set are different from the speakers in the training set and the development set. During training, the label order in the inference module is sorted in descending order according to speech energy.

In (inChannel, outChannel, kernel, stride)-format, for the encoder in the feature extraction module, 1D convolution has (1, 256, 40, 20)-kernel with no pooling. This corresponds to a frame length of 5 ms and a 2.5 ms shift. In the inference module, BLSTM layers have 3 layers with 256 nodes in each direction. Two LSTM-based decoders both run with 3 layers with 512 nodes. The dimension of the speaker and the direction embedding is 256. In the separation module, TCN runs with four convolution blocks and $R=8$ in each block. Transposed convolution runs with (256, 1, 40, 20)-kernel. For loss, $\lambda=5$. For SDNet, the input is raw-waveform, and it outputs raw-waveform.

\subsection{Baselines}

In our experiments, we build several baselines. SNet \cite{shi2019ones} acts as the baseline and is performed in the frequency domain. SNet-2ch represents the dual-channel version of SNet. We also build a dual-channel TasNet, named TasNet-2ch, whose channel differences are learned in an end-to-end manner. These models are trained with the same datasets as our models.

\subsection{Analysis of the proposed methods}

Learned from the experimental results in Table~\ref{tab:results1} and Table~\ref{tab:results2}, the proposed methods can effectively separate the mixed speech. In Table~\ref{tab:results1}, SNet is first transferred into the time domain as SNet-time. Compared with SNet, SNet-time achieves performance improvement, which attributes to the time-domain-based end-to-end training. SNet-time-2ch means the dual-channel SNet-time. Compared with SNet-time, SNet-time-2ch achieves a significant performance improvement. It means that the spatial information can be utilized by our network to improve the performance.

IAC is used to extract the differences between channels. The extracted features are only used in the inference module. The time-domain-based dual-channel model with IAC is named as SNet-time-2ch+IAC. As shown in Table~\ref{tab:results2}, the models with IAC have achieved performance improvement both on the anechoic and reverberant datasets.

When reverberation is added, performance is degraded. SDNet has achieved performance improvements both on anechoic and reverberant datasets. When separating the mixture, the speaker and direction can be inferred by SDNet. The inferred speaker and direction information is conducive to the selection of output. Our final model, SDNet, can achieve SDR improvements of 25.31 dB, 17.26 dB, and 21.56 dB on the anechoic WSJ0-2mix, WSJ0-3mix, and WSJ0-2\&3mix datasets and 10.64 dB, 8.55 dB, and 9.08 dB on the reverberant WSJ0-2mix, WSJ0-3mix, and WSJ0-2\&3mix datasets, respectively.

The performances of SNet-time and SNet-time-2ch are worse than the corresponding TasNets. This is due to the speaker mismatch between the training set and the test set, resulting in inaccurate speaker masks generated during the test. The direction inference mechanism in SDNet can effectively alleviate this problem. Reverberation has a negative impact on direction inference. SDNet performs similar to TasNet-2ch, but it does not need prior knowledge of the number of outputs.

\subsection{Inference accuracy of sound-source number}

The advantage of our proposed model is that it can dynamically estimate the number of sound sources. In the WSJ0-2mix and WSJ0-3mix, the number of speakers mixed in speech is fixed. We find that the models learn this pattern. In these experiments, the inference accuracies are close to 100\%. Therefore, we construct the WSJ0-2\&3mix dataset and perform experiments on this dataset. The experimental results are shown in Table~\ref{tab:results3}.

For comparison, SNet-time in Table~\ref{tab:results3} is performed on the same reverberant dataset but only on the reference channel. In Table~\ref{tab:results3}, compared with SNet-time, the inference accuracy of SNet-time-2ch has been greatly improved, which indicates that the spatial information is learned by the model and used to increases the discrimination of the sound sources. Compared with SNet-time-2ch, SNet-time-2ch+IAC can infer the number of sound sources more accurately. This shows that the extracted channel differences are beneficial to our system. SDNet achieves 89.73\%, which indicates the proposed method can make better use of the spatial information.

\begin{table}[t]
        \caption{\label{tab:results3} {\it Inferring accuracy of source numbers on reverberant WSJ0-2\&3mix dataset.}}
        \centerline{
          \begin{tabular}{c c}
            \toprule[1pt]
            {Model} & {Accuracy (\%)}    \\
            \hline
            \hline
              SNet-time          & $81.75$     \\
              SNet-time-2ch      & $85.11$     \\
              SNet-time-2ch+IAC  & $86.67$     \\
              SDNet              & $89.73$     \\
             \bottomrule[1pt]
          \end{tabular}
        }
\end{table}

\section{Conclusions}

We propose a time-domain-based speaker and direction-inferred dual-channel speech separation network, which first infers the speaker with direction and then integrates them as a source mask to separate the mixed speech. Experimental results show that SDNet effectively separates mixture under anechoic and reverberant conditions and deals with the problem of an unknown number of sources in the mixture and selection of outputs.

\section{Acknowledgments}
This work was done when Chenxing Li working as a visiting student at Columbia University. We thank Yi Luo and Cong Han for their helpful suggestions. Chenxing Li, Jiaming Xu, and Bo Xu were funded by a grant from the Major Project for New Generation
of AI (2018AAA0100400), and the Strategic
Priority Research Program of the Chinese Academy of Sciences
(XDB32070000).

\bibliographystyle{IEEEtran}
\bibliography{bibtex}

\end{document}